\documentstyle[aps,multicol,amssymb,psfig,float,times]{revtex}

\begin{document}


\title{BPS states in M-theory and twistorial constituents}

\author{Igor A. Bandos$^{\ast,\ddagger}$, Jos\'e A. de 
Azc\'arraga$^{\ast}$,
Jos\'e M. Izquierdo$^{\dagger}$ and Jerzy Lukierski$^{\ast ,\star}$}
\address{$^{\ast}$Departamento de F\'{\i}sica Te\'orica and IFIC,
 46100-Burjassot (Valencia), Spain}
\address{$^{\dagger}$Departamento de F\'{\i}sica Te\'orica,
 Facultad de Ciencias, 47011-Valladolid, Spain}
\address{$^{\ddagger}$Institute for Theoretical Physics, NSC KIPT, 
UA61108, Kharkov, Ukraine} 
\address{ $^\star$Institute for Theoretical 
Physics, pl. Maxa Borna 9, 50-204 Wroclaw, Poland}
\date{FTUV/01-0117, IFIC/01-03, January 17, 2001}

\maketitle

\begin{abstract}

We provide a complete algebraic description of the Bogomol'nyi-Prasad
Sommerfield (BPS) states in M-theory in 
terms of primary constituents that we call BPS preons. We argue 
that any BPS state preserving $k$ of the $32$ supersymmetries is a 
composite of $(32-k)$ BPS preons. In particular, the BPS states corresponding 
to the basic M2 and M5 branes are composed of $16$ BPS preons.
By extending the M-algebra to a generalized 
$D$=$11$ conformal superalgebra $osp(1|64)$ we relate the BPS preons 
with its fundamental representation, the 
$D$=$11$ supertwistors.

\end{abstract}

\begin{multicols}{2}

\narrowtext

1. {\it Introduction}. The dynamical description of the eleven-
dimensional
M-theory \cite{Wi.95,To.95}, which should unify all fundamental
interactions including gravity, is not known. It is characterized by a 
set of (conjectured) duality symmetries and by the low energy limit, 
$D$=$11$ supergravity \cite{Cr.79}. A relevant information is provided by the
$D$=$11$ Poincar\'e superalgebra recently called M-algebra \cite{1}
$$
    \{ Q_\alpha,Q_\beta\}=Z_{\alpha\beta}\quad ,
   \quad [Q_\alpha, Z_{\alpha\beta}]=0
    \; ,
$$
\begin{equation}
   Z_{\alpha\beta}\equiv\Gamma^\mu_{\alpha\beta}P_\mu +
      i\Gamma^{\mu\nu}_{\alpha\beta}Z_{\mu\nu}+
\Gamma^{\mu_1\dots\mu_5}_{\alpha\beta}Z_{\mu_1\dots\mu_5}\ ,
                                                           \label{1c}
\end{equation}
where $\Gamma^\mu_{\alpha\beta}=
{(\Gamma^\mu)_\alpha}^\gamma_. C_{\gamma\beta}$, etc, $Q_\alpha$ is a
$32$-component real Majorana spinor of supercharges, and the symmetric
$32\times 32$ generator $Z_{\alpha\beta}$  of tensorial 
(central with respect to $Q_\alpha$) charges extends the eleven momentum 
components $P_\mu$ to a set of $528=11+55+462$ generators. The additional 
$517$ charges characterize the basic M-branes. Assuming that the M-algebra 
is valid for all energies, the information about the spectrum of states in 
M-theory can be deduced from the representation theory of the algebra 
(\ref{1c}). Of special importance is the notion of the 
Bogomol'nyi-Prasad-Sommerfield (BPS) states. A BPS state 
$|k\rangle$ can be defined as an eigenstate with eigenvalue
$z_{\alpha\beta}$ of the `generalized momentum' generator,
$Z_{\alpha\beta}|k\rangle=z_{\alpha\beta}|k\rangle$, such that ${\rm det}\, 
z_{\alpha\beta}=0$:
 \begin{equation}
   \frac{k}{32}{\rm -}{\rm BPS \ state}:\, \{  {\rm
rank}\, z_{\alpha\beta}=32-k\ , 32>k\geq 1\}        \ .
\label{2} \end{equation}
Moreover \cite{To.97}, eq. (\ref{2}) implies that the BPS state  $|k\rangle$
preserves a fraction  $\nu=\frac{k}{32}$ of supersymmetries.

Without a knowledge of the fundamental dynamics of M-theory, it is
difficult to determine which representations of (\ref{1c}) are primary and
which are composite. A point of view based on the study of solitonic
solutions of $D$=$11$ supergravity \cite{Du.95,Hull.98} considers as the most
elementary ones the  $\frac{1}{2}$-BPS states describing the M2 
and M5 branes, the M9 brane and the M-KK6 brane ($D$=$11$ Kaluza-Klein 
monopole), as well as the M-wave (M0-brane). By considering superpositions of 
these $D$=$11$ elementary objects (intersecting branes and branes ending 
on branes) one can construct $\frac{k}{32}$-BPS $D$=$11$ supergravity 
solitons with $k\leq 16$ $(\nu\leq\frac{1}{2})$ \cite{To.97,Mo.00}. Despite
that exotic BPS states with $\nu=\frac{k}{32}>\frac{1}{2}$ can also be treated
algebraically as a kind of superposition of branes and antibranes
\cite{Ga.Hu.00,Mo.00}, solitonic solutions are known only
for BPS states that preserve a fraction 
$\nu\leq\frac{1}{2}$ of supersymmetries.

In this paper we propose another algebraic scheme aimed to 
describe the representations of M-algebra, with a different choice of primary 
and composite objects. Eq. (\ref{2}) suggests that the most elementary
component permitting to construct {\it all} BPS states as 
composites corresponds to tensorial charges with ${\rm rank}\, 
z_{\alpha\beta}=1$, or $k=31$. We shall call {\it BPS preons} the hypothetical 
objects carrying these `elementary values' of $z_{\alpha\beta}$. Thus, a 
BPS preon may be characterized by the following choice of central charges 
matrix
\begin{equation}
     z_{\alpha\beta}=\lambda_\alpha\lambda_\beta \ ,\quad
          \alpha,\beta=1,\dots, 32\ ,                         \label{3}
\end{equation} where $\lambda_\alpha$ is a real (Majorana) 
$SO(1,10)$ bosonic spinor. We notice that eq. (\ref{3}) can be looked at as an
extension of the Penrose formula (see {\it e.g.} \cite{Pe.72}) expressing a 
massless $D$=$4$ four-momentum as a bilinear of a Weyl spinor $\pi_A$ 
\begin{equation}
    p^\mu=\frac{1}{2}(\sigma^\mu)_{A{\dot B}}\pi^A{\bar\pi}^{\dot B}\ ,
    \quad  A,{\dot B}=1,2
                                     \ ,\label{4}
\end{equation}
to the case of $D$=$11$ generalized momenta (with abelian addition law).
Following the spirit of the twistor approach
\cite{Pe.72,Pe.77,Hu.79} we generalize eq. (\ref{3}) to
the general case of $\frac{k}{32}$-BPS states by
\begin{equation}
       z_{\alpha\beta}=\Sigma_{i=1}^n 
\lambda_\alpha^i\lambda_\beta^i\
       ,\quad
 n=32-k \ ,\quad 32>k\geq 1 \ ,
 \label{5}
\end{equation}
with  $n$ linearly independent $\lambda$'s.
Thus, the $\frac{k}{32}$-BPS state may be regarded as composed
 of $n=(32-k)$ BPS preons.

Group theory methods often compensate the lack of knowledge
about dynamical mechanisms. At this stage our aim is modest, and begins by 
pointing out an interesting structure of the representation theory of 
the M-algebra that arises when the tensorial charges are introduced as 
bilinears of spinors. Further, to conjecture the dynamics of BPS 
preons we may follow the
considerations in \cite{Ba.98,Ba.99} and enlarge the
M-algebra to the superconformal one $osp(1|64)$ 
\cite{vhol.pro.82,Bars}. In such a
dynamical picture a BPS preon (a fundamental $\frac{31}{32}$-BPS 
state) is described by a $D$=$11$ supertwistor
$(\lambda_\alpha,\omega^\alpha,\xi)$, with $\xi$ fermionic, {\it i.e.} by the
fundamental representation of $OSp(1|64)$.

We stress that our method provides a {\it complete} algebraic 
classification of all BPS states (note also that,
although we consider here the $D$=$11$
case, our approach applies to any $D$). In particular we shall
describe in the language of $n$ BPS preons the special choices
of $z_{\alpha\beta}$  describing M2 branes and M5 branes ($n=16$),
two orthogonal  $M$-branes ($n=24$) and the example \cite{Ga.Hu.00} of
an exotic BPS state with $\nu=\frac{3}{4}$. Due to the lack of $D$=$11$ 
solutions corresponding to exotic BPS states, one can conjecture that
branes that can be described in $D$=$11$ spacetime should have $n\geq 16$ in 
eq. (\ref{5}).

2. {\it Arbitrary BPS states as composites of BPS preons}. We
show now that any $\frac{k}{32}$-BPS state $|k\rangle$ can be characterized by
eq. (\ref{5}). $GL(32,\Bbb{R})$ is the maximal automorphism group of the algebra
(\ref{1c}) \cite{Ba.We.99,We.00,Ga.00}. As the matrix 
$z_{\alpha\beta}=\langle k|Z_{\alpha\beta}|k\rangle$ is
symmetric and positive-definite,
         $z_{\alpha\beta}x^\alpha 
x^\beta=2\Sigma_{l=1}^{32}|\langle k|x^\alpha{\hat
         Q}_\alpha|l\rangle|^2\geq 0$,                                    
it can be diagonalized by a $GL(32,\Bbb{R})$ matrix
${G^\alpha_{.}}_\beta\; ,$
 \begin{equation}
       z_{\alpha\beta}={G^\gamma_{.}}_\alpha z_{\gamma\delta}^{(0)}
{G^\delta_.}_\beta \ .                                  \label{7}
\end{equation}
Moreover the diagonal matrix $z_{\gamma\delta}^{(0)}$ can be chosen as
follows 
\begin{equation}
    z_{\gamma\delta}^{(0)}={\rm diag}(1,\mathop{\dots}\limits^n,1,0,
    \mathop{\dots}\limits^k,0)    \ .\label{8}
\end{equation}
Thus, we see that eq.
(\ref{7}) can be written in the form (\ref{5}) provided that
\begin{equation}
  \lambda^i_\alpha={G^i_.}_\alpha \ , \quad i=1,\dots ,n=32-k  \
  .\label{9}
\end{equation}
In the new basis $Q_\alpha^{(0)}$ defined by
$Q_\alpha^{(0)}={(G^{-1})^\beta_.}_\alpha Q_\beta$, the M-algebra (\ref{1c}) 
diagonalizes on BPS states so that
\begin{eqnarray}
         \{ Q_i^{(0)},Q_j^{(0)}\}|k\rangle &=& \delta_{ij} |k\rangle\ , \nonumber\\
        \{ Q_i^{(0)},Q_r^{(0)}\}|k\rangle &=& \{ Q_r^{(0)},Q_s^{(0)}\}|k\rangle=0\ ,
                                     \label{10}
\end{eqnarray}
$r,s=n+1,\dots,32$. Hence, the set of $32$ supercharges
$Q_\alpha^{(0)}=(Q_i^{(0)},Q_r^{(0)})$ acting on the BPS state $|k\rangle$
splits into $k$ generators $Q_r^{(0)}$ of supersymmetries 
preserving the
BPS state ({\it i.e.} one can put consistently $Q_r^{(0)}|k\rangle=0$), and
$n=32-k$ generators $Q_i^{(0)}$ which describe the set of broken
supersymmetries. Summarizing, the eigenvalues $z_{\alpha\beta}$ 
of the tensorial charges that characterize a BPS state preserving $k<32$
supersymmetries may be expressed by eq. (\ref{5}) in terms of {\ } $32-k$  
Majorana $Spin(1,10)$ bosonic spinors $\lambda_\alpha^i$. A  BPS preon 
state with $z_{\alpha\beta}=\lambda_\alpha\lambda_\beta$
preserves $k=31$ supersymmetries and $\nu=\frac{k}{32}$ BPS states are
composed of $n=32-k$ preons.  

Eq. (\ref{5}) implies that the tensorial charges are preserved
under $O(n)$ transformations of the $n$ spinors $\lambda_\alpha^i$:
\begin{equation}
     \lambda_\alpha^i{'}={O^i}_j\lambda_\alpha^j\ , \quad
                                     O^TO={\bf 1}_n\ ,
                            \quad i=1,\dots ,n\; .             \label{12}
\end{equation}
The $SO(n)$=$SO(32-k)$ rotations constitute an
internal symmetry of the $\frac{k}{32}$-BPS states. For 
the $\frac{1}{2}$-BPS states associated with the fundamental M2 and 
M5 branes, in particular, $SO(n)$=$SO(16)$. This group corresponds in 
$D$=$11$ to the unitary 
internal symmetry $U(n)$ in $D$=$4$ that was introduced in the framework of 
twistor theory for $n$-twistor composite systems \cite{Pe.77,Hu.79}. In 
our scheme we see that the dimension $n$ of $SO(n)$ is equal to the number of 
broken supersymmetries.

3. {\it Physical $\frac{1}{2}$ BPS states and their superpositions.}
The fundamental BPS preons have all the  bosonic charges
$p_\mu \propto \lambda C\Gamma_\mu \lambda$,
$ z_{\mu\nu} \propto\lambda C\Gamma_{\mu\nu} \lambda$,
$ z_{\mu_1\ldots \mu_5 }\propto\lambda C\Gamma_{\mu_1\ldots 
\mu_5 } \lambda$ nonvanishing. We show now in our framework that the 
M2 and M5 $1\over 2$--BPS states can be composed out of 
16 BPS preons.

Let us consider the M2 brane, what implies $z_{\mu_1\ldots \mu_5 
}=0$. Thus, eq. (\ref{5}) acquires the form
\begin{equation}\label{3.1}
z_{\alpha \beta}= p_\mu \Gamma^\mu_{\alpha\beta} + z_{\mu\nu }
i\Gamma^{\mu\nu}_{\alpha\beta}= \Sigma_{i=1}^n \lambda^i_{\alpha}
\lambda^i_{\beta}\; ,
\end{equation}
where, at this stage, we do not fix $n$. In the rest frame 
of a BPS massive state, $p_\mu= m (1,0, \ldots ,0)$.
As we are dealing with the M2 brane, but not with the M9 brane,
we shall assume that in this frame  $z_{\mu\nu}$ has only space-like
components $z_{\mu\nu}= \delta^I_{[\mu}\delta_{\nu]}^J z_{IJ}$. 
In particular we may choose the space slice of the M2 worldvolume in the
$\{12\}$ plane, $z_{\mu\nu}=
\delta^1_{[\mu}\delta_{\nu]}^2 z$ where $z=2z_{12}$.
Using the $Spin(1,2)\otimes Spin(8)$ covariant splitting of $D$=$11$ 
spinors and $\Gamma$ matrices (see {\it e.g.} \cite{BZ}),
\begin{equation}
\label{3.2} \lambda^i_{\alpha}= \left(\matrix{\lambda^i_{a
q} \cr
\lambda^{ia}_{\dot q}\cr}  \right)\ ,\quad a=1,2 \; ,\quad q,{\dot
q}=1,\dots ,8 \; , \end{equation}
$i\Gamma_{12}=\pmatrix{I_{16} &0 \cr 0 &-I_{16}}$, 
$\gamma^0_{ab}=\delta_{ab}$, eq. 
(\ref{3.1}) can be written as $$
\hbox{ \bf M2}: \qquad \matrix{1 & 2 & - &- &- &- &- &- &- &- &- & 
\cr} $$
\begin{eqnarray}\label{3.3}
z_{\alpha \beta}&=& \left(\matrix{(m+z)
\delta_{ab}\delta_{qp} & 0 \cr
                   0 & (m-z)\delta^{ab}\delta_{{\dot q}{\dot p}} \cr}
                   \right)
\nonumber\\
&=&
\Sigma_{i=1}^n\left(\matrix{\lambda^i_{a q}\lambda^i_{b p} &
\lambda^i_{a q}\lambda^{ib}_{\dot p}
\cr \lambda^{ia}_{\dot q} \lambda^i_{b p} & \lambda^{ia}_{ \dot q}
          \lambda^{ib}_{\dot p}\cr}\right)\; .
\end{eqnarray}
The matrix $z_{\alpha \beta}$ has either rank $32$ (when $m \not=
\pm z$), or rank $16$ (when $m= \pm z$). Assuming $z>0$ we conclude
that the M2 brane BPS state appears when $m= z$ and that preserves 
1/2 of the target supersymmetries. In this case Eq. (\ref{3.3}) implies
\begin{equation}
\label{3.4} \Sigma_{i=1}^n(\lambda^i_{a q}\lambda^i_{b
p})= 2z\delta_{ab}\delta_{qp}\ ,
\end{equation}
\begin{equation}
\label{3.5}
\Sigma_{i=1}^n(\lambda^{ia}_{\dot q}\lambda^i_{b
p})=0=
\Sigma_{i=1}^n(\lambda^{ia}_{\dot q}\lambda^{ib}_{\dot p})
\ . \end{equation}
Eq. (\ref{3.4}) has a solution only if $n\geq
16$. Moreover, as rank$\, z_{\alpha\beta}$=$16$ for $m=z$, we need just $16$ 
BPS preons, described by $\lambda_{aq}^i\in {\Bbb{R}}^+\otimes O(16)$; from 
(\ref{3.5}) $\lambda^{ia}_{\dot p}=0$ follows. Using the $O(16)$ symmetry 
(eq. (\ref{5})) we see that in a special frame the spinors 
$\lambda_{\alpha}^i$ ($i=1,\dots ,16$)
satisfying eqs. (\ref{3.4}-\ref{3.5}) and 
describing a M2-brane BPS state may be written as $\lambda^i_{(\alpha )}$,
\begin{equation}\label{3.6} \hbox{ \bf M2}\ (m=z): \quad \quad \quad 
{}
\lambda^i_{(\alpha )}= \left(\matrix{\sqrt{2z} \delta^i_{a q} \cr 0 \cr}
\right)\ .
                                     \end{equation}
To obtain the set of $\lambda_{\alpha}^i$ in an arbitrary frame we
perform a Lorentz rotation of (\ref{3.6}) by means of a $Spin(1,10)$ matrix
\begin{eqnarray}\label{3.7}
 v_{\alpha}{}^{(\beta )} &\equiv &
\left(v_{\alpha}{}^{aq},  v_{\alpha a \dot{q}}\right)
\in  Spin(1,10)  \ , \nonumber\\
 a &= &1,2\; , \; q=1,...,8\; ,\  \dot{q}=1,...,8\; ,
\end{eqnarray}
and we get
 \begin{equation}\label{3.8}
\hbox{ \bf M2}: \qquad {}\qquad {}
\lambda^i_{\alpha}=  v_{\alpha}{}^{(\beta )}
\lambda^i_{(\beta )} =\sqrt{2z} v_{\alpha}{}^{aq} \delta^i_{a q}\; .
\end{equation}
Note that only one of the two 
$[Spin(1,2) \otimes Spin(8)]$-covariant $32\times 16$ blocks 
({\it cf.} \cite{BZ}) of the spinorial Lorentz frame matrix 
(\ref{3.7}), $v_{\alpha}{}^{aq}$, enters in eq. 
(\ref{3.8}) \cite{2}. 

For a M5-brane BPS state corresponding to the vanishing worldvolume gauge
field the central charges matrix is
\begin{equation}\label{3.10}
z_{\alpha \beta}= p_\mu \Gamma^\mu_{\alpha\beta}+  
z_{\mu_1\ldots \mu_5 }
\Gamma^{\mu_1\ldots \mu_5  }_{\alpha\beta}= \Sigma_{i=1}^{16}
\lambda^i_{\alpha} \lambda_{\beta}^i \, .
\end{equation}
By considerations analogous to those of M2-brane case
one finds that the $16$ BPS preons needed for such a BPS state of 
a M5-brane are associated with
\begin{equation}
\label{3.11} \hbox{ \bf M5}: \qquad 
\lambda^i_{\alpha}=  v_{\alpha}{}^{(\beta )} \lambda^i_{(\beta )}
=\sqrt{2z} v_{\alpha}{}^{aq} \delta^i_{a q} \; ,\end{equation} $$ a=1,...,4
\quad (Spin(1,5)) \qquad q=1,...,4  \quad (Spin(5)), $$
where $z=m= 5!z_{1...5}>0$, and $v_{\alpha}{}^{aq}$
is the $16 \times 32$ $[Spin(1,5)\otimes Spin(5)]$-covariant 
block of the Lorentz frame matrix $v_{\alpha}{}^{(\beta )}$.
Thus, the M2 and M5 BPS states are
described by a highly constrained set of $D$=11 spinors
$\lambda_{\alpha}^i$ (since $v_{\alpha}{}^{(\beta)}$ in 
eq. (\ref{3.8}) as well as in eq. (\ref{3.11}) belongs
to $Spin(1,10)$).

To describe a superposition of M-branes preserving 
$\nu ={k\over 32} < {1\over 2}$ supersymmetry one needs  
$n> 16$ BPS preons. In particular, for the system of two 
orthogonal M2-branes 

\hfill $ \matrix{1 & 2 & - &- &- &- &- &- &- &- &- & \cr} $

\noindent {\bf M2 $\otimes$ M2}
 
\hfill   $ \matrix{- & - & 3 & 4&- &- &- &- &- &- &- & \cr} $

\noindent with equal positive charges  $z_{12}=z_{34}=m/4$ 
in the rest frame we get 
\begin{equation}
\label{M2M2}
z_{\alpha\beta}= m\pmatrix{ \gamma^{0}_{ab}\delta_{qp} + 
{\cal P}^{(-)}_{aq ~bp} & 0 \cr
 0 & {\cal P}^{(-)}{}^{ab}_{\dot{q}\dot{p}} }=\Sigma_{i=1}^{24}\lambda_\alpha^i
 \lambda_\beta^i  \; , 
\end{equation}
where ${\cal P}^{(-)}{}^{ab}_{\dot{q}\dot{p}}= {1\over 2} 
({\gamma}^{0~ab}\delta_{\dot{q}\dot{p}}- \varepsilon^{ab} 
\gamma^1_{q\dot{q}}\gamma^2_{q\dot{p}} )$ is the orthogonal 
projector, $\epsilon^{12}=1$,
$\gamma^{0,1,2}_{ab},\gamma^{0ab},\dots ,$ are $SO(1,2)$ gamma matrices and
$\gamma^1_{q{\dot q}},\dots ,\gamma^8_{q{\dot q}}$ are the $8\times 8$ $SO(8)$
Pauli matrices. Thus, {\rm rank}$\, z_{\alpha\beta}= 24$ and we need 
$24$ BPS preons which can be characterized by 
($\hat{i}=1,...,16;~\tilde{i}=1,...,8$, $i=1,...,24$)
\begin{equation}
\label{LM2M2}
\lambda_{\alpha}^i= \left(
\pmatrix{\hat{\lambda}^{~\hat{i}}_{aq}\cr 0 \cr },  
\pmatrix{0 \cr  \tilde{\lambda}^{a\tilde{i}}_{\dot{q}}\cr }
\right),  \end{equation}
where the $\hat{\lambda}^{~\hat{i}}_{aq}$ and the  
$\tilde{\lambda}^{a\tilde{i}}_{\dot{q}}$ are constrained by 
\begin{eqnarray}
\label{cL1}
\Sigma_{\hat{i}=1}^{16} \hat{\lambda}^{~\hat{i}}_{aq}
\hat{\lambda}^{~\hat{i}}_{bp}&=&m(\gamma^{0}_{ab}\delta_{qp} + 
{\cal P}^{(-)}{}_{aq ~bp})\; ,
\nonumber \\ 
\Sigma_{\tilde{i}=1}^{8}  \tilde{\lambda}^{a\tilde{i}}_{\dot{q}} 
\tilde{\lambda}^{b\tilde{i}}_{\dot{p}}
&=& m{\cal P}^{(-)}{}^{ab}_{\dot{q}\dot{p}}\; .  
\end{eqnarray}
In an arbitrary frame $\lambda_{\alpha}^i= (
v_{\alpha}{}^{aq}\hat{\lambda}^{~\hat{i}}_{aq},  
v_{\alpha a \dot{q}} \tilde{\lambda}^{a\tilde{i}}_{\dot{q}})$,  where 
$
v_{\alpha}{}^{aq}, v_{\alpha a \dot{q}}$ are the spinor
frame variables (\ref{3.7}).  

The ${1\over 4}$-BPS state (\ref{M2M2}) as well as many other 
$\nu ={k\over 32}< {1\over 2}$ 
BPS states are described by solitonic solutions of the 
$D$=11 supergravity \cite{Du.95}. 
Our approach allows to consider as well the `exotic' BPS states with 
$\nu > {1\over 2}$. They are described by  $n=32-k< 16$ BPS
preons. For instance, the massive ${3\over 4}$-BPS state 
in \cite{Ga.Hu.00}, with $z_{12}=-m/2$, 
$z_{2345}=-z_{26789\# }=\mp m/5!$ in the rest frame,  
is characterized by 
\begin{equation}
\label{z3/4}
z_{\alpha\beta}=4m\pmatrix{ 0
& 0 \cr
 0 & {\cal P}^{(\pm )}{}^{ab}_{\dot{q}\dot{p}}  } =
 \Sigma_{i=1}^{8}\lambda_\alpha^i
 \lambda_\beta^i \; , 
\end{equation}
where now ${\cal P}^{(\pm )}{}^{ab}_{\dot{q}\dot{p}}= {1\over 2} 
({\gamma}^{0~ab}\delta_{\dot{q}\dot{p}} \pm \gamma^{2~ab} 
{\gamma}^{1234}_{\dot{q}\dot{p}} )$ are orthogonal projectors  
(${\cal P}^{(+)}+ {\cal P}^{(-)}= {\gamma}^{0}\otimes I$, 
${\gamma}^{1234}_{\dot{q}\dot{p}}\equiv 
\gamma^1_{q\dot{q}}\gamma^2_{q\dot{r}} \gamma^3_{p\dot{r}}
\gamma^4_{p\dot{p}}= - {\gamma}^{5678}_{\dot{q}\dot{p}}$). 
Thus, {\rm rank}$\, z_{\alpha\beta}= 8$ and one concludes that the BPS 
state preserves $\nu = 3/4$ of supersymmetry  \cite{Ga.Hu.00}. 
In an arbitrary frame 
$z_{\alpha\beta}= 4m v_{\alpha a \dot{q}} 
{\cal P}^{(\pm )}{}^{ab}_{\dot{q}\dot{p}} v_{\beta b \dot{q}}$  
and the 
${3\over 4}$-BPS state can be described by 
$8$ BPS preons 
\begin{equation}
\label{L=v3/4}
\lambda_{\alpha}^{a\tilde{i}}= v_{\alpha a \dot{q}} 
\tilde{\lambda}^{a\tilde{i}}_{\dot{q}}\; , 
\quad \tilde{i}=1,...,8\; ,
\end{equation}
where the $8$ $\tilde{\lambda}^{a\tilde{i}}_{\dot{q}}$ are 
constrained by $\Sigma_{\tilde{i}=1}^{8}  
\tilde{\lambda}^{a\tilde{i}}_{\dot{q}} 
\tilde{\lambda}^{b\tilde{i}}_{\dot{p}}
= 4m 
{\cal P}^{(\pm )}{}^{ab}_{\dot{q}\dot{p}}$
and $v_{\alpha}{}^{aq}, v_{\alpha a \dot{q}}$ are the 
spinorial Lorentz frame variables of eq. (\ref{3.7}).

There are no solitonic solutions for the exotic BPS states known. 
Moreover, general $\kappa$-symmetry arguments \cite{BKO} and the 
study of the simplest supersymmetric field theories \cite{Ga.00} 
indicate that, probably, such solitonic solutions do not exist in the 
standard ({\it i.e.} unenlarged) $D$=$11$ spacetime. This suggests that 
only composites of $n\geq 16$ BPS preons can be described in a $D$=$11$ 
standard spacetime framework.

4. {\it BPS preons, enlarged superspaces and OSp$(1|64)$ supertwistors}. 
It seems natural to assume that a
dynamical realization of exotic states requires a new geometric framework,
going beyond the standard $D$=$11$ spacetime \cite{3}.
The most straightforward idea is to treat 
all tensorial charges as generalized momenta in a large conjugate 
space of $528$ dimensions. The simplest supersymmetric dynamics in 
$D$=$4$ superspace --Brink-Schwarz massless superparticle-- 
can be extended to such a
large space by two different ways of generalizing the mass-shell
condition:

i) The  Sp$(32)$-invariant generalization 
$z_{\alpha\beta}C^{\beta\gamma}z_{\gamma\delta}=0$ of $p^2=0$  
({\it cf.} \cite{Se.97}), where the Sp$(32)$ metric $C$
is the antisymmetric $D$=$11$ Majorana charge-conjugation matrix.

ii) The general, less restrictive, $GL(32,\Bbb{R})$-invariant condition
(see {\it e.g.} \cite{Ga.00}) ${\rm det}\, z_{\alpha\beta}=0$,
characterizing all $\frac{k}{32}$-BPS states with $1\leq k< 32$.

To introduce a
dynamical scheme for the proposed BPS preons of M-algebra one can develop a new
spinorial geometry by doubling the $D$=$11$ Lorenz spinors to 
introduce a $D$=$11$ twistor $T_A=(\lambda_\alpha,\omega^\alpha)$ 
$(A=1,\dots,64)$ satisfying a generalized Penrose incidence equation
\begin{equation}
        \omega^\alpha=x^{\alpha\beta}\lambda_\beta\ ,\quad
           x^{\alpha\beta}=x^{\beta\alpha}\ ,              \label{24}
\end{equation}
 where
\begin{equation}
      x^{\alpha\beta}=x^\mu\Gamma_\mu^{\alpha\beta}+y^{\mu\nu}
i\Gamma_{\mu\nu}^{\alpha\beta}
+y^{\mu_1\dots\mu_5}\Gamma_{\mu_1\dots\mu_5}^{\alpha\beta}  
\label{25}
\end{equation}
describes the $528=11+55+462$ coordinates dual to the $P_\mu$,
$Z_{\mu\nu}$, $Z_{\mu_1\dots\mu_5}$ generalized momenta.
In a supersymmetric theory, eq. (\ref{24}) has to be supplemented by 
({\it cf.} \cite{Fe.78,Ba.98,Ba.99})
\begin{equation}
\label{30bis}
       \xi=\theta^\alpha\lambda_\alpha\ .
\end{equation}
Then, ${\cal T}_{\cal A}=(T_A,\xi)$ defines a supertwistor,
which is the fundamental representation of the generalized $D$=$11$ 
conformal superalgebra $osp(1|64)$. In such a framework the basic 
geometry is described by the $D$=$11$  supertwistors ($T_{A},\xi$) which
we propose to interprete as BPS preon phase space coordinates. 
Indeed, using eqs. (\ref{24}) and (\ref{5}) one obtains a
relation (modulo an exterior derivative) between the canonical 
Liouville one-forms describing the symplectic structure 
in the enlarged spacetime (\ref{25}) and the $D$=$11$ twistor 
space coordinates,
\begin{equation}
       z_{\alpha\beta}dx^{\alpha\beta}=\Sigma^n_{i=1}\lambda^i_\alpha
        \lambda^i_{\beta} dx^{\alpha\beta}=-2\Sigma^n_{i=1}
          \omega^{\alpha i}d\lambda_\alpha^i \; ,            \label{31}
\end{equation}
a relation that can be supersymmetrized \cite{Ba.98,Ba.99}.
For non--BPS states, for which
${\rm det}\, z_{\alpha\beta}\neq 0$, one needs the maximal 
number, $32$, of BPS preons described by $32$ supertwistors 
${\cal T}_{\cal A}^i=(\lambda_\alpha^i,\omega^{\alpha i},\xi^i)$. 
Using $32$ copies of the eqs. (\ref{24}), (\ref{30bis}) one can express all 
enlarged superspace coordinates $(x^{\alpha\beta},\theta^\alpha)$
as composites of spinorial preonic coordinates as follows:
\begin{equation}
          x^{\alpha\beta}=\Sigma_{i=1}^{32}\omega^{\alpha i}
            (\lambda^{-1})^\beta_i\; ,\; 
           \theta^\alpha=\Sigma_{i=1}^{32}\xi^{ i}
            (\lambda^{-1})^\alpha_i\, .                    \label{32}
\end{equation}
If we diminish the number of BPS preons, the geometry becomes gradually more
spinorial and detached from a spacetime framework. In particular, the
most elementary constituent of M-theory matter in the present
approach, a single BPS preon, is 
described by the purely spinorial geometry of a single supertwistor.

\vspace{0.2cm}

{\it Acknowledgments}. This work has been partially supported by 
the DGICYT research grant PB 96-0756, the Ministerio de Educaci\'on 
y Cultura (I.B.), the Generalitat Valenciana 
and KBN grant 5 P03B 05620 (J.L.) and the Junta de Castilla y
Le\'on (research grant C02/199). We thank J. Sim\'on for helpful
discussions.

\vspace{0.2cm}

 {\it E-mails}: bandos@ific.uv.es,
j.a.de.azcarraga@ific.uv.es,\\
 izquierd@fta.uva.es, lukier@ift.uni.wroc.pl


\end{multicols}

\begin{thebibliography}{99}
\bibitem{1}
We shall only consider here the conventional form of
$D$=$11$ M-algebra \cite{To.97,vhol.pro.82}, without additional fermionic 
(spinorial) charges. For its general form see \cite{Se.96}. For the 
topological origin of `central' charges and their role in the 
partial breaking of supersymmetry see \cite{Az.89}; for a general
discussion of extended supersymmetry algebras and extended 
superspaces see \cite{Ch.00}.
\bibitem{2}
This block is used in the Lorentz harmonics formulation of the 
worldvolume action of the M2--brane \cite{BZ}
to build the projector of the irreducible
$\kappa$--symmetry $ {1\over 2}{(1+
\bar{\Gamma})_\alpha}^\beta_. = v_{\alpha}{}^{aq} v_{aq}{}^{\beta} $.
Thus, for the M2--brane BPS state, eq. (\ref{3.1}) implies\\
$\langle 16_{M2}|\{Q_{\alpha},Q_{\beta} \} |16_{M2}\rangle = 
z_{\alpha \beta} =
p_\mu\Gamma^\mu_{\alpha\beta}+  z_{\mu\nu} 
i\Gamma^{\mu\nu}_{\alpha\beta}$\\
$\quad \quad =  2zv_{\alpha}{}^{aq} v_{\beta}{}^{bq}
\gamma^0_{ab} = 2z{1\over 2}((1+ 
\bar{\Gamma})\Gamma^{(0)}C)_{\alpha \beta}\; ,$\\
(where the $SO(1,2)$ $\gamma^0_{ab}=\delta_{ab}$ and $\Gamma^{(0)}$
is defined in \cite{BZ}) which
 reflects the deep relation \cite{BKO,ST} between the
supersymmetry of a BPS state and the $\kappa$--symmetry of the 
worldvolume action of the corresponding superbrane.
\end{thebibliography}
\end{document}